\documentclass[12pt, preprint]{aastex}

\slugcomment{accepted for publication in \emph{Icarus}}

\shorttitle{Meteorology of Jupiter's Equatorial Hot Spots and Plumes from \emph{Cassini}}
\shortauthors{Choi et al.}

\begin{document}

\title{Meteorology of Jupiter's Equatorial Hot Spots and Plumes from \emph{Cassini}}

\author{David S. Choi}
\affil{ORAU / NASA Goddard Space Flight Center, Greenbelt, MD 20771}
\email{david.s.choi@nasa.gov}
\email{david@davidchoi.org}

\author{Adam P. Showman}
\affil{Department of Planetary Sciences, The University of Arizona,
				Tucson, AZ 85721}
				
\author{Ashwin R. Vasavada}
\affil{Jet Propulsion Laboratory, California Institute of Technology, Pasadena, CA 91011}

\and

\author{Amy A. Simon-Miller}
\affil{NASA Goddard Space Flight Center, Greenbelt, MD 20771 (U.S.A.)}

\begin{abstract}
We present an updated analysis of Jupiter's equatorial meteorology from \emph{Cassini} observations. For two months preceding the spacecraft's closest approach, the Imaging Science Subsystem (ISS) onboard regularly imaged the atmosphere. We created time-lapse movies from this period in order to analyze the dynamics of equatorial hot spots and their interactions with adjacent latitudes. Hot spots are relatively cloud-free regions that emit strongly at 5 microns; improved knowledge of these features is crucial for fully understanding Galileo probe measurements taken during its descent through one. Hot spots are quasi-stable, rectangular dark areas on visible-wavelength images, with defined eastern edges that sharply contrast with surrounding clouds, but diffuse western edges serving as nebulous boundaries with adjacent equatorial plumes. Hot spots exhibit significant variations in size and shape over timescales of days and weeks. Some of these changes correspond with passing vortex systems from adjacent latitudes interacting with hot spots. Strong anticyclonic gyres present to the south and southeast of the dark areas appear to circulate into hot spots. Impressive, bright white plumes occupy spaces in between hot spots. Compact cirrus-like 'scooter' clouds flow rapidly through the plumes before disappearing within the dark areas. These clouds travel at 150-200 m/s, much faster than the 100 m/s hot spot and plume drift speed. This raises the possibility that the scooter clouds may be more illustrative of the actual jet stream speed at these latitudes. Most previously published zonal wind profiles represent the drift speed of the hot spots at their latitude from pattern matching of the entire longitudinal image strip. If a downward branch of an equatorially-trapped Rossby waves controls the overall appearance of hot spots, however, the westward phase velocity of the wave leads to underestimates of the true jet stream speed. 
\end{abstract}

\keywords{Jupiter --- Jupiter, atmosphere --- Atmospheres, dynamics --- Meteorology}

\section{Introduction}

Unique atmospheric phenomena reside in Jupiter's Equatorial Zone. For example, hot spots are compact, quasi-periodic areas of strong infrared emission confined between 5$^{\circ}$N and 10$^{\circ}$N \citep{Rogers95}. These hot spots are bright when observing Jupiter's thermal emission in the 5 $\mu$m infrared window, but are dark when observing reflected sunlight in visible or near-infrared wavelengths. These observations suggest that hot spots are areas of reduced cloud opacity, allowing transmission of infrared radiation from deeper, warmer layers of Jupiter's troposphere. Analysis of \emph{Galileo} data by \citet{Banfield98} show that stratospheric and upper tropospheric layers are present above hot spots, but the ammonia cloud layer just above 1 bar pressure ubiquitous throughout most of the atmosphere is not present. Further analysis by \citet{Simon-Miller01} concluded that the vertical structure above Jovian hot spots was unique, with no substantial clouds at pressures 1 bar or higher. 

In-situ observations made by the \emph{Galileo} entry probe \citep{Young03} as it descended through the southern fringes of a hot spot strengthened the importance of understanding Jovian hot spot meteorology. Given the quasi-periodic spacing of the hot spots, the leading theory for these features is that a downwelling portion of an equatorial planetary wave is responsible for their formation and maintenance \citep{Allison90, Ortiz98, Showman00, Friedson05}. The numerical models of \citet{Showman00} broadly reproduce observations, lending more credence to the equatorial wave theory as a mechanism responsible for these phenomena. However, certain details, such as the flow pattern around hot spots, mutual hot spot interactions, and factors influencing hot spot morphology, have yet to be examined in detail by simulations.

\citet{Vasavada98} analyzed \emph{Galileo} observations of a single hot spot and reported winds entering a hot spot at its southwest, suggesting some degree of convergent flow at a hot spot. However, this study could not constrain how winds flowed away from a hot spot. Numerical simulations by \citet{Showman00} reproduced the observed flow entering a hot spot at its southwest and suggested that air flows out of a hotspot at its southeast. In contrast, \citet{Garcia-Melendo11} recently reported that flow entering a hot spot at its southwest was present in only two of nine hot spots from an analysis of \emph{Cassini} imagery. Therefore, it is unclear if the \emph{Galileo} observations indicate a pattern shared by all hot spots.

Other previous studies illuminate a broad picture of Jupiter's equatorial meteorology. Historical records and observations examined by \citet{Rogers95} demonstrate that 10--12 hot spots (i.e. ``dark projections'') have been consistently present on Jupiter throughout the 20th century. \citet{Ortiz98} examined a more recent 3-year infrared dataset leading up to the \emph{Galileo} probe entry and concluded that the observed hot spots are confined to a narrow latitude band centered at $\sim$7$^{\circ}$ N and that at least 10 or 11 were present during their observation period, roughly evenly spaced in longitude. Furthermore, their study found that although an individual hot spot can experience growth, decline, merger, splitting, or reshaping over days to weeks, the overall pattern of hot spots is stable for months to years. A more recent study by \citet{Arregi06} that examined hot spots in visible observations stated that hot spots exhibited a dispersion relationship between drift speed and wave number, which was dominant between 8-12.

Equatorial plumes are immense white cloud streaks located in between hot spots and embedded within the Equatorial Zone \citep{Reese76}. \citet{Stoker86} proposed that the plumes were thunderstorm anvil-like structures generated from moist convection within a storm center. However, this theory encounters difficulties because lightning would presumably be present \citep{Gierasch00}, and has not yet been observed. Furthermore, only a few plume heads are bright white and suggestive of convection \citep{Rogers95}. 

An alternative explanation is that plumes represent the upward branch of an equatorial planetary wave \citep{Allison90, Showman00}, where systematic lifting throughout the entire plume area is responsible for cloud condensation, and where plumes and any storms present are independent of one another. \citet{Showman05} constrained the wave model by demonstrating ammonia depletion in the plumes, rendering advection from depth via convection unlikely. However, the observed sub-saturation in ammonia does not eliminate the possibility of thunderstorms within them. \citet{Simon-Miller00} detected possible water ice spectral signatures on Jupiter from \emph{Voyager} data concentrated near the latitudes of the hotspots and plumes, suggestive of convection in or near these areas, but also consistent with wave models that allow but do not require storm activity. Furthermore, \citet{Baines02} spectroscopically detected discrete ammonia ice clouds in Jupiter's atmosphere, and most of them were located within a small latitudinal band just to the south of hot spots. Their proximity and consistent relative location to hot spots further strengthened the case for a wave controlling these features. Overall, however, deficiencies in our knowledge regarding the local meteorology of plumes prevent a complete understanding of the relationship between plumes, hot spots, and the surrounding atmosphere.

In this paper, we examine an extensive \emph{Cassini} data set composed as the spacecraft approached Jupiter. We focus on the dynamics of Jupiter's equatorial region by distilling the typical motions occurring within equatorial hot spots, plumes, and their surroundings. Our results are largely descriptive, though we include quantitative results when available. We also examine correlations between changes in ambient winds with changes in hot spot morphology. 

\section{Description of Data Set}

We compiled data from \emph{Cassini}'s Imaging Science Subsystem (ISS) \citep{Porco04}. As the spacecraft approached Jupiter, the ISS regularly acquired global, multispectral imaging data for just over two months. Typical imaging sequences using a particular filter consisted of observations taken approximately every 100 minutes throughout a single Jovian day. The spacecraft usually acquired data on consecutive Jovian days, with occasional gap days. After mid-November 2000, as \emph{Cassini} drew closer, a series of images arranged in a 2x2 grid was necessary to capture the entire planet for a single observation. The spatial resolution of the images improved gradually from $\sim$500 km pixel$^{-1}$ to $\sim$130 km pixel$^{-1}$. Sub-spacecraft and sub-solar latitudes for all images were both $\sim$3$^{\circ}$N, with the solar phase angle ranging from 5$^{\circ}$ to 20$^{\circ}$. 

The ISS observed Jupiter using a filter set at ultraviolet, visible, and near-infrared wavelengths, covering both continuum and gaseous absorption bands. In this study, we use images observed with the CB2 filter (750 nm central wavelength), a near-infrared continuum region of the spectrum ideal for dynamical and morphological studies due to its minimal Rayleigh scattering and gaseous absorption. Cloud structures and morphological features are seen in high contrast, rendering CB2 data ideal for cloud feature tracking.

After compiling all available CB2 data, we navigated and mapped the images using a rectangular (simple cylindrical) projection. We utilized the latest SPICE navigational kernels for the best available pointing accuracy. 

\section{Methodology}

The regular observations enabled the creation of image pairs covering the same area on Jupiter but separated by the $\sim$100-minute time interval. Such pairs are fit for analysis using automated cloud feature tracking \citep{Choi07}. We processed all available pairs with our software using a 3$^{\circ}$ square correlation window for initial guidance and a 1$^{\circ}$ square window for fine measurement tuning. After processing, we manually removed spurious results through visual inspection of the wind map and identification of measurements incongruous relative to its surroundings. Unfortunately, full-longitudinal maps of the wind field are not possible in most cases because of the aforementioned observation gaps, especially during the initial portion of the overall imaging campaign. Data transmission difficulties and other technical problems also caused other coverage gaps. 

We used measurement tools integrated into the SAOImage DS9 software suite \citep{Joye03} to mark locations of hot spots and plumes, and to measure their sizes. Unfortunately, there is unavoidable subjectivity when measuring hot spot sizes because their complex morphology typically prevents a straightforward assessment of a hot spot's longitudinal extent. Therefore, an individual measurement of a hot spot's size includes some level of uncertainty. However, broader long-term trends evident from a collection of measurements are more robust.

Apart from image pairs, we constructed time-lapse movies (see online supplementary information) showing the morphological evolution of hot spots, plumes, and surrounding areas. Our movies have a dynamic frame-of-reference that matches the drift speed of specific features; for the hot spots, this speed is typically 103 m s$^{-1}$ eastward. 

\section{Results}

\subsection{Hot Spot Locations and Size}

\begin{figure}[htbp]
  \centering
  \includegraphics[keepaspectratio=true, totalheight=5.5in]{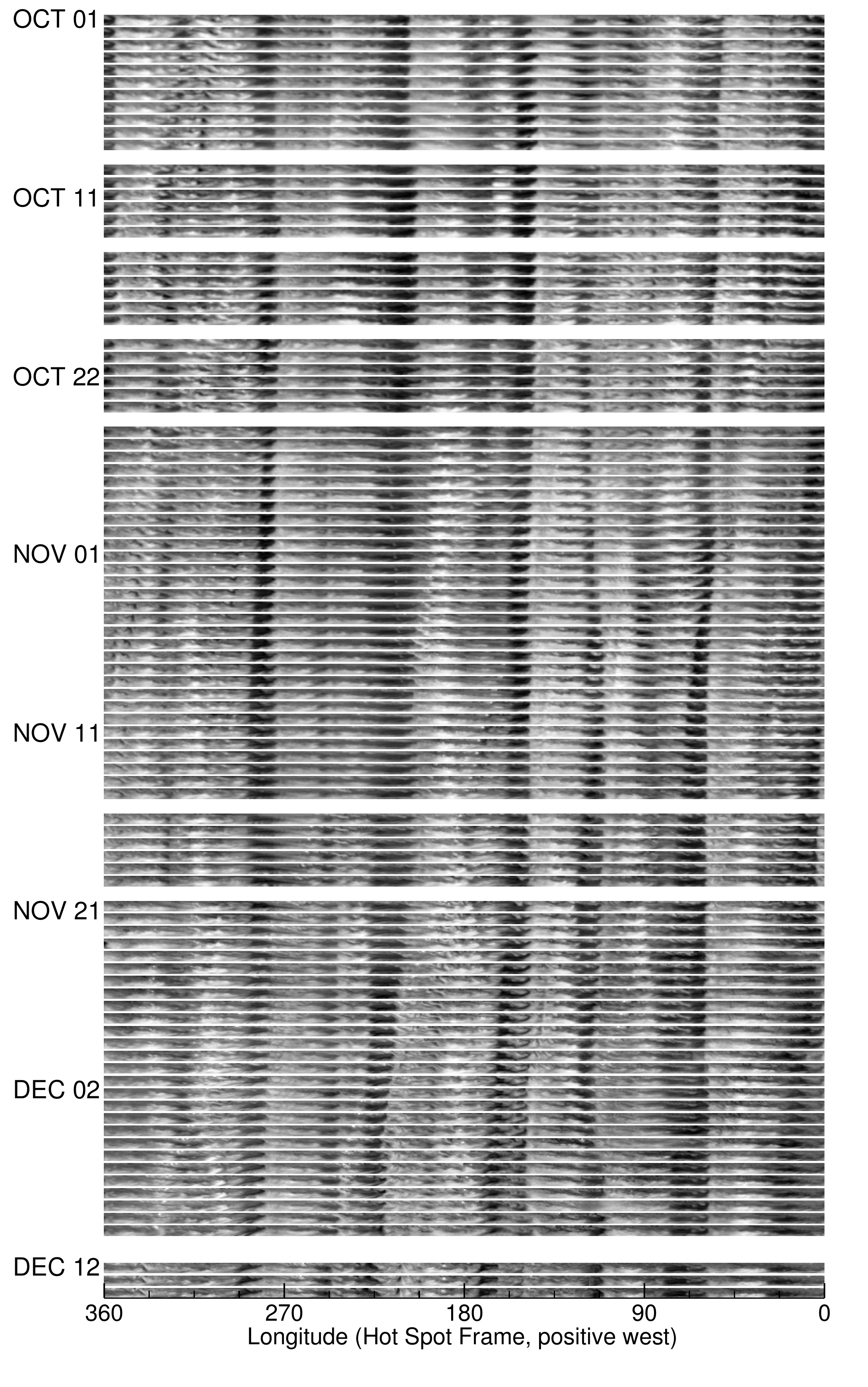}
  \caption[Hot Spot Mosaic Strips vs. Time]{
    \label{Figure: hotspot_mosaicvstime}
    Temporal evolution of Jupiter's North Equatorial Belt illustrated using numerous \emph{Cassini} ISS latitudinal strips centered at 7.5$^{\circ}$N. Each strip spans 360$^{\circ}$ longitude, 5$^{\circ}$ latitude and is comprised of images observed during a single Jovian day. Most strips shown in this figure represent consecutive Jovian days, though some do not due to observational constraints. Each strip is longitudinally shifted to a reference frame traveling east at 103 m s$^{-1}$, the propagation speed of the hot spots.
    }
\end{figure}

Figure \ref{Figure: hotspot_mosaicvstime} is a time-series of \emph{Cassini} image strips throughout the observation period fixed to a reference frame tied to the propagation speed of the hot spots and plumes. (An annotated version of Figure \ref{Figure: hotspot_mosaicvstime} is available online as supplementary material. These annotations mark the locations and times of events referenced in the figures and text.) Although small-scale morphological details are not visible in Figure \ref{Figure: hotspot_mosaicvstime}, it provides an excellent overview of the dynamic meteorology occurring near Jupiter's equator. Within the two-month span, various examples emerge of hot spot stability, migration, growth, decline, split, and merger. In addition, various dark spots and bright clouds appear and disappear on timescales of days and weeks. These features deviate from a vertical in Figure \ref{Figure: hotspot_mosaicvstime}, demonstrating that they propagate at different speeds relative to the hot spots. As they traverse across the more prominent hot spots and plumes, these minor spots and clouds may be upwelling and downwelling components of secondary equatorial wave modes confined within this latitudinal band. 

\begin{figure}[htbp]
  \centering
  \includegraphics[keepaspectratio=true, totalheight=5.5in]{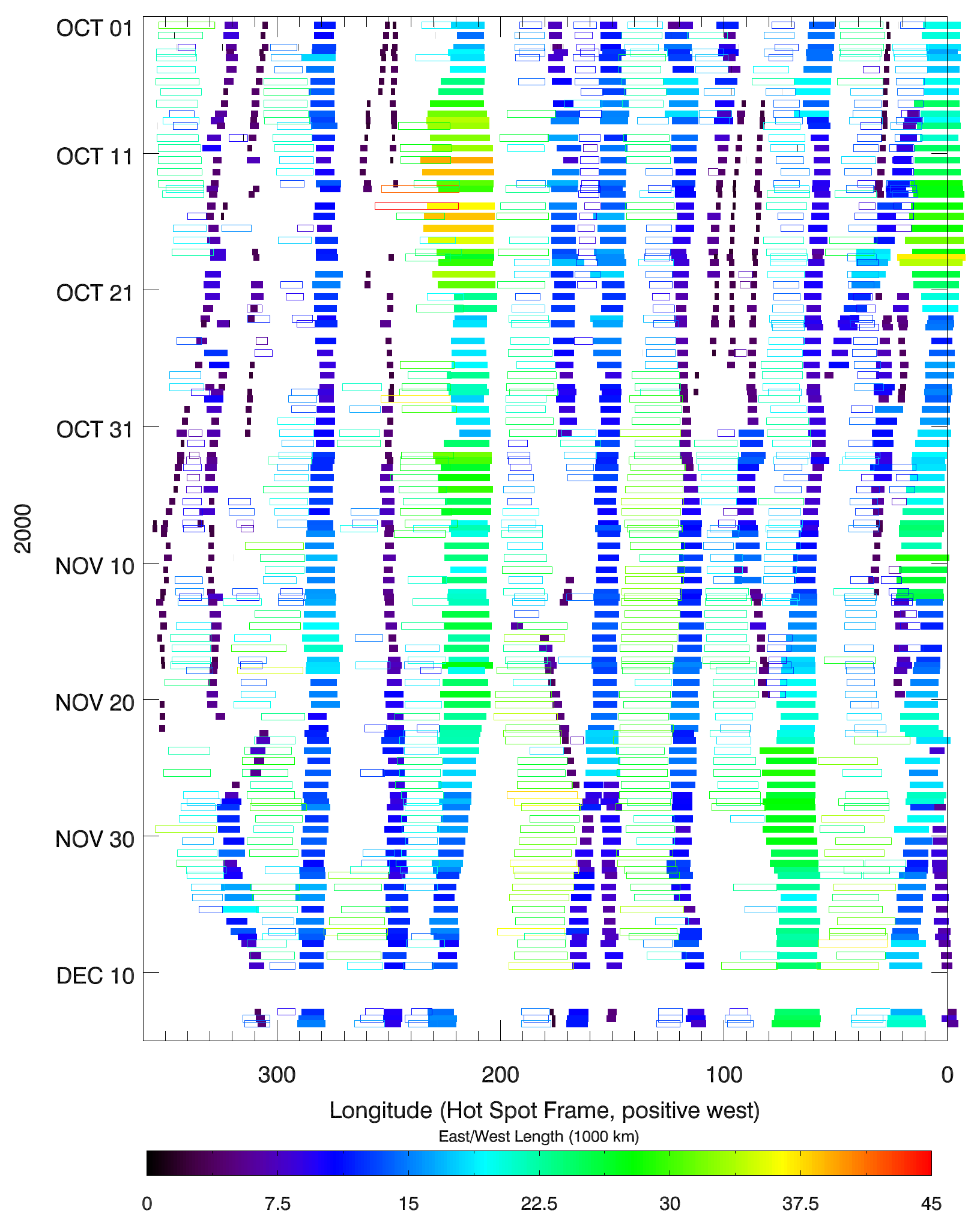}
  \caption[Hot Spot Locations vs. Time]{
    \label{Figure: hotspot_locvstime}
    Chart of hot spot and plume locations and sizes during the \emph{Cassini} observation campaign. Filled rectangles symbolize dark areas in the North Equatorial Belt, presumably representative of hot spots, whereas unfilled rectangles denote bright plumes. Each rectangle's size in the figure is a faithful representation of its observed dimensions. Color (in the online version of this figure) also scales with feature size. This figure shifts longitudes to a reference frame that is traveling eastward at 103 m s$^{-1}$, the apparent dominant propagation speed of the hot spots. 
    }
\end{figure}

We draw and define sharp boundaries for the equatorial features at 7.5$^{\circ}$N in Figure \ref{Figure: hotspot_locvstime}, as it illustrates the locations and sizes of dark areas and plumes observed within the \emph{Cassini} data set with filled-in and empty rectangles, respectively. The minor dark spots traveling at different phase speeds are arguably more visible here compared with Figure \ref{Figure: hotspot_mosaicvstime}. Throughout the observational period, most hot spots experience dynamic fluctuations in their size and shape, in addition to splitting and merging events. One notable exception is a hot spot located at roughly 290--280$^{\circ}$ longitude\footnote{Longitudes in this paper refer to an artificial reference frame that travels at the propagation speed of the hot spots, instead of System III.}. This hot spot experiences slight size variations but lacks significant transformative events. The relative position of hot spots with each other is relatively stable, suggesting that the overall propagation speed of hot spots is stable on weekly to monthly timescales. Furthermore, the overall stability of multiple hot spots and plumes with respect to one another supports the theory that these features are the result of a planetary-scale wave acting across the equatorial atmosphere, as opposed to being discrete, independent phenomena. 

Because hot spot dark areas are substantial cloud clearings in the ammonia cloud deck, obscurement of these areas by high-altitude clouds can cause the disappearance and subsequent reappearance of several small dark areas illustrated in Figure \ref{Figure: hotspot_locvstime} (e.g. $\sim$250$^{\circ}$ W longitude between Oct 11-21, see Figure \ref{Figure: suppl_fig1} at the end of this manuscript) in visible observations. Similarly, plume-like features can obscure dark areas (e.g. $\sim$20$^{\circ}$ W longitude between Nov 10-20), leading to the appearance of hot spot splitting. If a planetary wave is present, supplementary radiative transfer studies at these areas will constrain the uppermost altitude in which the downwelling branch of this wave influences on the troposphere. 

\begin{figure}[htbp]
  \centering
  \includegraphics[keepaspectratio=true, width=5.5in]{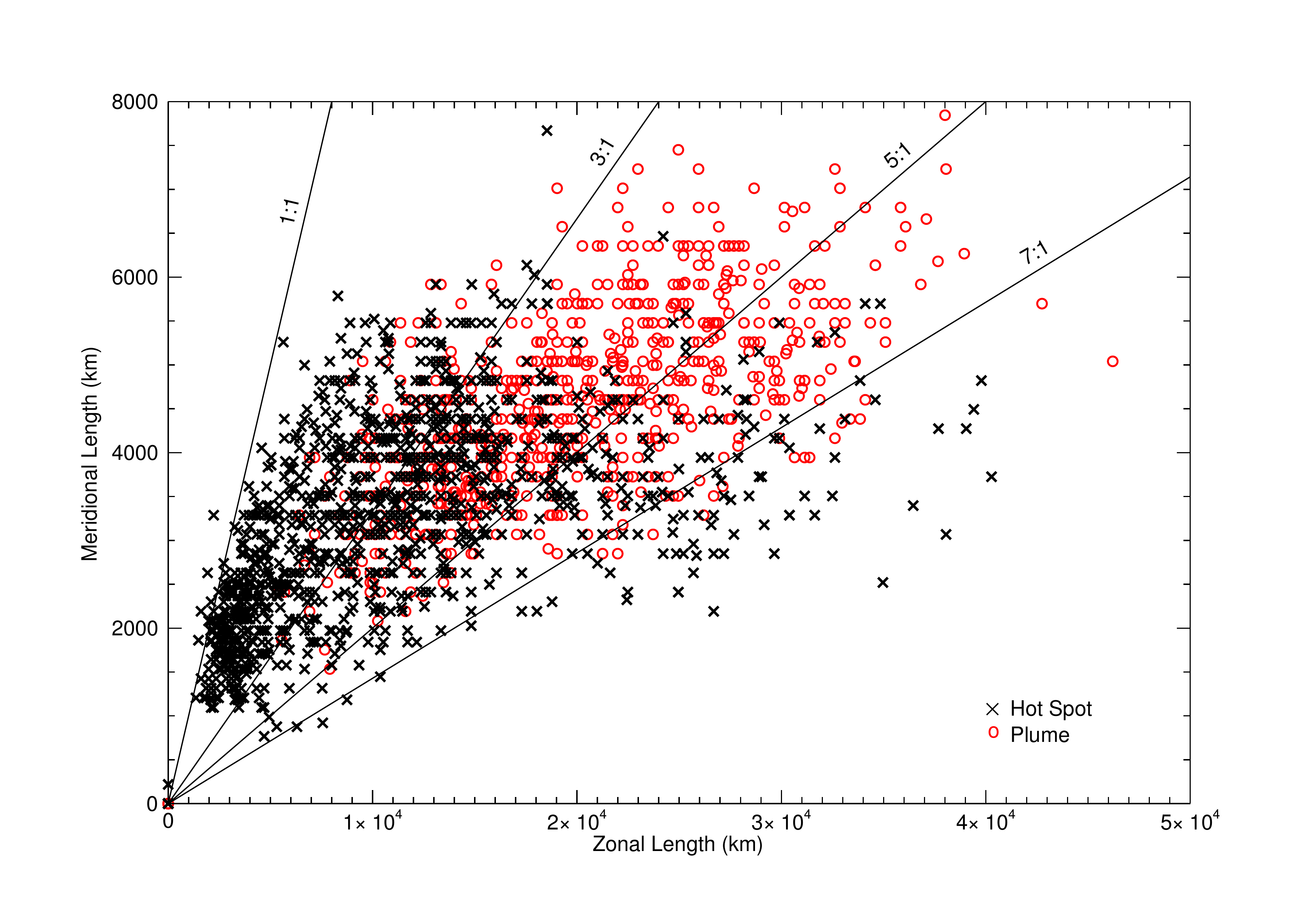}
  \caption[Hot Spot and Plume Sizes]{
    \label{Figure: hotspot_sizes}
    Scatter plot of hot spot and equatorial plume sizes in our data set. The plot includes all measurements from every observation. Lines denoting various aspect ratios for the zonal to meridional lengths are shown in the figure.
    }
\end{figure}

Figure \ref{Figure: hotspot_sizes} demonstrates that hot spots generally have aspect ratios between $\sim$1:1 and 7:1 (longitude-to-latitude). Overall, hot spots are generally narrower than plumes, as the median zonal length of a hot spot in our data set is just under 10,000 km, whereas the median meridional length is just over 3,000 km. In contrast, the median zonal length of equatorial plumes is nearly double that of the hot spots, though meridional lengths are similar to those of the hot spots. Strong zonal flows at their northern and southern boundaries confine the maximum meridional sizes of both features to be no more than 8000 km ($\sim$8$^{\circ}$). The size imbalance exhibited by these features suggests an asymmetry in the downwelling and upwelling branches of the planetary wave, if one is present. Alternatively, a large fraction of this wave may be above the condensation level for tropospheric ammonia.

\subsection{Hot Spot Morphology}
\subsubsection{The Quasi-stable, Time-Averaged State}

When observed in the near-infrared continuum (750 nm), a typical hot spot is a compact, quasi-rectangular dark area exhibiting some of the lowest brightness measurements for the entire planet at these wavelengths, suggesting either areas of low cloud opacity or inherently dark chromophores coloring the clouds. Infrared observations of elevated brightness temperatures within these areas confirm reduced cloud opacity \citep{Orton98} and render the presence of inherently dark chromophores as unlikely. The interactions of a hot spot with its surroundings, particularly along its boundaries, produce dark filamentary structures suggestive of dark material, but are likely low-opacity regions experiencing turbulence and shear with little horizontal mixing.

\begin{figure}[htbp]
  \centering
  \includegraphics[keepaspectratio=true, totalheight=2.5in]{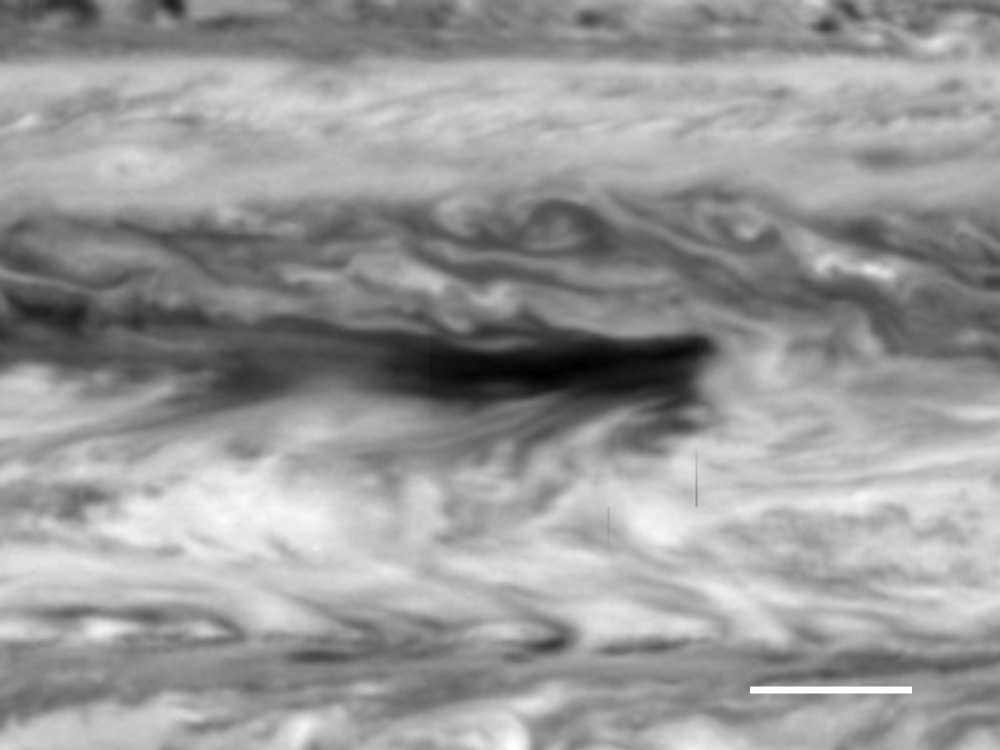}
  \caption[Jovian Hot Spot Observed at 750 nm]{
    \label{Figure: ideal_hotspot}
    Typical morphology of a Jovian hot spot, observed using the \emph{Cassini} ISS CB2 (750 nm) filter. North is up in this figure. Note the sharp transition at the hot spot's eastern edge, whereas the western edge gradually transitions to areas of increased cloud opacity. The scale bar at lower right denotes 10,000 km.
    }
\end{figure}

We illustrate the typical morphology of a hot spot in Figure \ref{Figure: ideal_hotspot}. Hot spots typically exhibit a sharply defined eastern edge demarcating the dark hot spot region from surrounding cloudy areas. This edge appears similar to a shock front. From a frame of reference traveling with the hot spots, incoming cloud material (typically from the north) appears to divert around the hot spot as it maintains its position. Occasionally, however, the edge retreats or experiences significant changes in shape when encountering particularly prominent storms and bright clouds from the North Equatorial Belt. After these interactions, the eastern edge can advance towards the east, restoring the hot spot's size before the encounter. However, other encounters result in splitting or quasi-permanent size changes, which we will discuss in section \ref{Section: interactions}.

In contrast, the western edge of a hot spot is not as clearly defined. Some portions of the western edge represent a gray transition region between hot spot and plume, whereas other portions clearly distinguish these features. The complexities associated with the western edge complicate measurements of hot spot sizes, add some uncertainty to these measurements, and raise questions over whether a hot spot is growing or dividing itself into two. The boundary between the North Equatorial Belt and a hotspot can also be latitudinally irregular at certain points, with arms or spits of cloud material extending into a hotspot at isolated areas.

\subsubsection{Inherent Changes}

\begin{figure}[htbp]
  \centering
  \includegraphics[keepaspectratio=true, totalheight=2.5in]{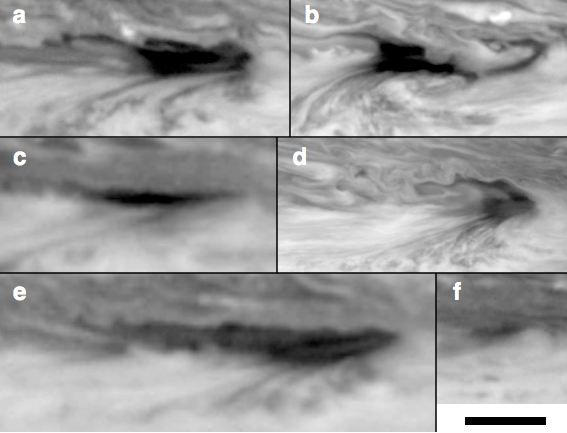}
  \caption[Jovian Hot Spot Observed at 750 nm]{
    \label{Figure: anomalous_hotspots}
    Various anomalous morphologies of Jovian hot spots, observed using the \emph{Cassini} ISS CB2 (750 nm) filter. North is up and the scale bar at lower right denotes 10,000 km for all panels in this figure. (a) Typical hot spot morphology. (b) The same hot spot in panel (a), nearly 2 weeks later, experiencing a splitting event. (c) Latitudinal compression of a hot spot. (d) Hot spot in a compact, square state. (e) Maximum longitudinal span of a hot spot during our observational data set. (f) Precursor dark spot, where a more stable and typical hot spot morphology emerged several weeks later. See Supplementary Figure 1 for the timing and locations of these anomalous morphologies in our overall dataset. 
    }
\end{figure}

The instantaneous state of a hot spot is somewhat unpredictable and subject to a variety of changes in its morphology as a result of its interactions with ambient winds and surrounding phenomena. Several `stable' hot spots (the set of hot spots present throughout most or all of the observational period that maintain their position relatively well) only experience minor variations in size and position. Other stable hot spots, however, are highly dynamic in shape and size, growing and shrinking by up to a factor of 2. Figure \ref{Figure: anomalous_hotspots} illustrates examples of these anomalous, yet relatively ephemeral, morphologies. Hot spots can split (Fig. \ref{Figure: anomalous_hotspots}a-b), where one portion remains at its original propagation speed, while the other portion travels westward relative to the original spot at a modified propagation speed. Under the planetary wave model, this behavior suggests the formation and simultaneous existence of a new wave mode concurrent with the antecedent and dominant mode controlling the stable hot spots. However, the temporal limitations of our data set prevent knowledge of whether or not the split portion traveling at a different propagation speed does so for an extended time period. Furthermore, the portion that remains at the original propagation speed is typically much smaller and deformed, and may not be stable long-term. Because these splitting events typically occur in conjunction with passing storms in adjacent latitudes, we describe additional details regarding hot spot splitting events in Section \ref{Section: interactions}. 

Other changes to a hot spot's general morphology include latitudinal compression (Fig. \ref{Figure: anomalous_hotspots}c), where its latitudinal extent diminishes for an extended period of time. It is not clear if this decrease is real or is an illusion caused by cloud bands from the north, obscuring the dark areas. At other times, hot spots can dramatically shrink to become square and boxy (Fig. \ref{Figure: anomalous_hotspots}d), or longitudinally extended (Fig. \ref{Figure: anomalous_hotspots}e). Boxy, compact states typically last for shorter periods than rectangular extended states. Finally, precursor dark spots (Fig. \ref{Figure: anomalous_hotspots}f) are small dark features that are more latitudinally compressed than mature dark spots, and travel at the same drift speed as the stable hot spot group. These spots episodically disappear and re-emerge from surrounding clouds before achieving a more mature morphology after several weeks of instability. A fundamental issue shared among all anomalous states is the degree to which these observed changes are the result of inherent variability in the planetary waves driving equatorial meteorology, or if they are the result of obscuring cloud layers at different heights and/or from other latitudes. Both factors likely play a role, and further study is necessary to constrain their relative importance. 

\subsubsection{Interactions with Adjacent Latitudes}
\label{Section: interactions}

Storm systems and vortices residing in latitudes both north and south of the hot spots influence their morphology. One particular system present during the \emph{Cassini} observations is an alternating system of anti-cyclonic and cyclonic vortices moving in tandem residing north of the hot spots and plumes in the North Equatorial Belt. This pattern resembles the White Ovals throughout most of the 20th century \citep{Youssef03} and is perhaps a von K\'arm\'an vortex street. In a frame of reference traveling eastward with the hot spots, this vortex triplet travels rapidly westward because it is embedded in the westward jet stream within the North Equatorial Belt. The influence of the vortices as they pass near hot spots is evident in the rapidly changing sizes and morphologies of hot spots after interaction. Typically, turbulent cloud patterns accompanying the vortices or associated with bright cloud tops generated as the vortices travel (presumably thunderstorms) directly interact with hot spots as these cloud patterns travel southward into the hot spots.

\begin{figure}[htbp]
  \centering
  \includegraphics[keepaspectratio=true, totalheight=5.5in]{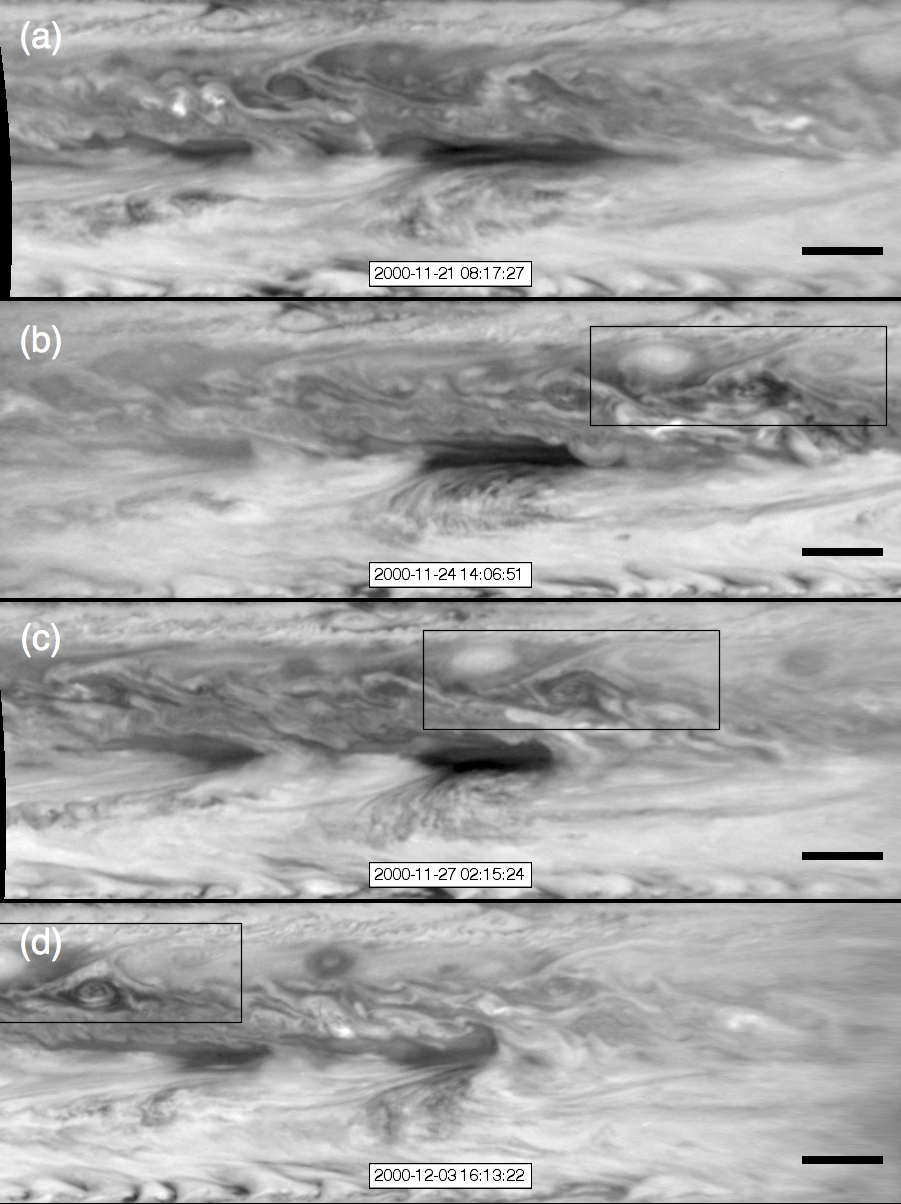}
  \caption[Jovian Hot Spot Interaction with Vortex Street]{
    \label{Figure: hotspot_vortex_interaction}
    Time series of \emph{Cassini} ISS CB2 images depicting a Jovian hot spot and the effects of a passing vortex street (outlined in a rectangle in panels b-d in figure) to its north. (a) Initial state: hot spot is rectangular and has a relatively large longitude-to-latitude aspect ratio. (b) Vortices approach: seemingly turbulent cloud patterns encroach upon the dark spot at its northeastern quadrant. (c) Peak interaction: the vortices and associated clouds buffet against the hot spot, shaping its boundary into a curve. (d) Aftermath: the hot spot's longitudinal extent is diminished. The scale bar at the lower right in each panel denotes 10,000 km.
    }
\end{figure}

Figure \ref{Figure: hotspot_vortex_interaction} illustrates the interaction between the vortex street and a hot spot. In Figure \ref{Figure: hotspot_vortex_interaction}a, a hot spot is initially in an elongated rectangular state. As the vortices approach from the east, a cloud front intrudes southward. The intrusion, seen in Figure \ref{Figure: hotspot_vortex_interaction}b, dramatically decreases the apparent size of the hot spot. This hot spot behavior contrasts with behavior observed during an interaction with an advancing cloud front from the north two weeks prior. Earlier, the hot spot maintained its apparent size against this cloud front, with the front wrapping around the northeastern quadrant of the hot spot. In essence, the visual appearance was as if the hot spots were a solid body advancing against the incoming fluid flow. Differences in the energetics of the advancing front likely explains these contrasting behaviors, as the cloud front that overtakes the hot spot is associated with the vortex triplet complex.

In figure \ref{Figure: hotspot_vortex_interaction}c, the vortex triplet makes their closest approach to the hot spot. Here, the associated cloud complex overwhelms the hot spot, substantially decreasing in size. In the time-lapse animations, dark areas shed away from the hot spot and advect away to the north and south, and then to the west. Bright spots (thunderstorms?) also appear during closest approach: one of these is visible to the west of the hot spot in figure \ref{Figure: hotspot_vortex_interaction}d. These thunderstorms shed turbulent ribbons of cloud material that interact with the hot spot. At the conclusion of the encounter, a dark area emerges in the area of the hot spot's former maximum eastern extent. While this could represent a new hot spot, it is more likely that hazes or clouds that had occupied this area are vacating, uncovering a dark area below. At the conclusion of the \emph{Cassini} observations, the hot spot's morphology is greatly modified and is at roughly half of its previous size.

Other hot spot encounters with the vortex complex, however, result in behaviors ranging from little to no effect to the splitting of a hot spot. This implies that the strength of the wave controlling a particular hot spot relative to the energetics of the incoming cloud features influences the hot spot's resulting behaviors during encounters. In one example of one end of the behavioral spectrum, a hot spot's morphology and size do not change substantially. Only minor intrusions of cloud material from the north shroud the hot spot, along with minor latitudinal compression. Future numerical modeling simulations can investigate the role of the anticyclonic gyre to the southeast of a hot spot, and how its strength affects the maintenance of a hot spot's morphology during and after its encounter with incoming material.  

\begin{figure}[htbp]
  \centering
  \includegraphics[keepaspectratio=true, totalheight=5.5in]{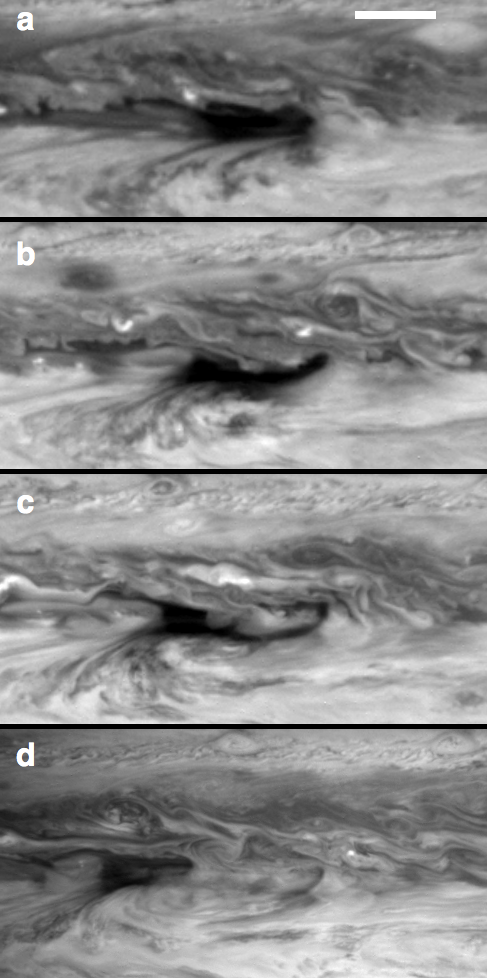}
  \caption[Jovian Hot Spot Splitting Event]{
    \label{Figure: hotspot_split}
    A hot spot splitting event observed using the \emph{Cassini} ISS CB2 (750 nm) filter. North is up and the scale bar at upper right denotes 10,000 km for all panels in this figure. (a) The hot spot before the split. (b) Clouds begin advancing southward into the hot spot. (c) Clouds continue to advance southward and expand into the hot spot. The western portion of the hotspot begins to drift to the west. (d) Towards the end of our data set, the anticyclonic gyre has migrated to the west along with one split portion of former hot spot, whereas the other split portion has slightly drifted to the east and appears diminished.
    }
\end{figure}

At the opposite end of the behavioral spectrum, the passage of a vortex street appears to trigger a hot spot splitting event. Figure \ref{Figure: hotspot_split} illustrates this particular case, where the encounter proceeds with incursions of cloud material into the hot spot from the North Equatorial Belt. This intrusion of material perhaps acts as a catalyst for splitting, because the western portion of the newly split hot spot begins to drift west relative to the main hot spot propagation speed, indicating that the phase speed of this portion has increased westward. Furthermore, an anticyclonic gyre accompanies the split western portion of the hot spot, continuing to circulate while also traveling slightly northward. Meanwhile, the drift speed of the eastern portion remains unchanged, but becomes completely shrouded by clouds from the North Equatorial Belt advecting southward. By the end of the observation period, it is unknown if the eastern portion of the hot spot persists. What factors or parameters that ultimately control hot spot behavior during these encounters is a subject for future studies.

Another system that affected hot spots during the two-month observation period of \emph{Cassini}'s approach was the South Equatorial Disturbance (SED) \citep{Simon-Miller12}, which was active in the South Equatorial Belt to the south of the hot spots. The SED is a singular anticyclonic storm traveling westward relative to the southern branch of Jupiter's eastward-flowing equatorial jet stream. The SED primarily disturbs dark chevron-shaped cloud features embedded within this jet stream; as chevrons encounter the SED, they are displaced southward and merge with the storm, but afterwards emerge or reform to the east of the SED \citep{Simon-Miller12}.

Though the SED directly influences meteorology at its resident latitude, the system may also influence hot spot morphology to its north. As the SED passes to the south of one prominent hot spot, the hot spot appears to split into two distinct portions. Much like splitting events triggered by the passage of vortex streets, clouds from the North Equatorial Belt advect southward turbulently into the hot spot latitudes. Some background darkness, presumably at deeper altitudes, remains visible through these clouds. However, the resulting behavior of the split hot spot portions implies fundamental changes to these features. The eastern portion of the split hot spot is much smaller and slightly increases its propagation speed eastward, advancing a modest amount. This portion may represent a different hot spot wave mode. The western portion of this split hot spot extends westward at the same speed as the SED, implying that the SED exerts some influence on the portion's growth. Unfortunately, this was the only passage of the SED with a mature hot spot in the data set, and it is unknown whether similar effects occur when the SED passes other prominent hot spots.

A peculiar feature is a dark spot within the hot spot latitudes located to the north and west of the SED. (This spot is notably seen in Movie \#4 of the supplemental online material, and is annotated in Supplementary Figure 1.) In the first part of the movie, this dark spot acts like a companion to the SED as it travels westward at the same speed as the SED. As this ``companion spot'' travels westward, it interacts with immature hot spots, temporarily increasing their size and darkening these spots. The ``companion spot'' also experiences significant fluctuations in its own morphology as it undergoes several cycles of growth, disappearance, and re-emergence. The physical mechanisms driving the morphological changes in the ``companion'' are unclear without further examination and analysis of multi-spectral data. It is possible that clouds from surrounding latitudes obscure the spot, causing an apparent morphological change. Finally, as the ``companion'' approaches the prominent hot spot that later splits (as discussed in the preceding paragraph), it slows its advance, eventually becoming stationary while also disappearing. Our observations of the ``companion'' appearing to pass through smaller, immature hot spots relatively unaffected imply wave interaction behavior that is linear or weakly non-linear, similar to mutual soliton interactions. The small stature and irregular morphology of these spots suggest that they may be weak perturbations, which may play a role in their resulting mutual interactions. In contrast, previous studies of ground-based images by \citet{Rogers95} and \citet{Arregi06} report mergers of prominent hot spots, suggesting that linear or weakly non-linear wave dynamics cannot provide a full explanation for the observed behavior across the entire spectrum of hot spot types.

\subsection{Hot Spot Dynamics}

The descent of the \emph{Galileo} entry probe through a hot spot motivated research into its meteorology, starting with basic questions regarding the atmospheric dynamics around a hot spot. \citet{Vasavada98} first reported 30--50 m s$^{-1}$ winds flowing southwest-to-northeast near the southwestern quadrant of a hot spot in their analysis of \emph{Galileo} data. This flow appeared to be associated with an anticyclonically rotating cloud located to the hot spot's southeast that could not be confirmed during their study. The overall flow pattern observed by \citet{Vasavada98}, however, resembled the steady-state flow pattern in numerical simulations by \citet{Showman00}.

Our time-lapse movies of the equatorial region uniquely reveal atmospheric motions, prioritizing a broader observational baseline throughout a period preceding \emph{Cassini}'s closest approach in exchange for reduced (though steadily improving) spatial resolution. Examination of these movies uncovered complex flow patterns between all meteorological features located in the equator and surrounding latitudes. We also applied our automated cloud feature tracker to the data, though with mixed results. Displacements of features in image pairs separated by $\sim$1 hour at reduced spatial resolution were insufficient to resolve meridional motion perceived when observing time-lapse footage. Analysis of image pairs separated by $\sim$10 hours somewhat resolved this issue but introduced new complexities originating from feature deformation beyond recognition to independent dynamics occurring at separate altitudes. In these cases, we relied on our observations from the time-lapse footage to complete our analysis of the dynamics. 

\begin{figure}[htbp]
  \centering
  \includegraphics[keepaspectratio=true, width=5.5in]{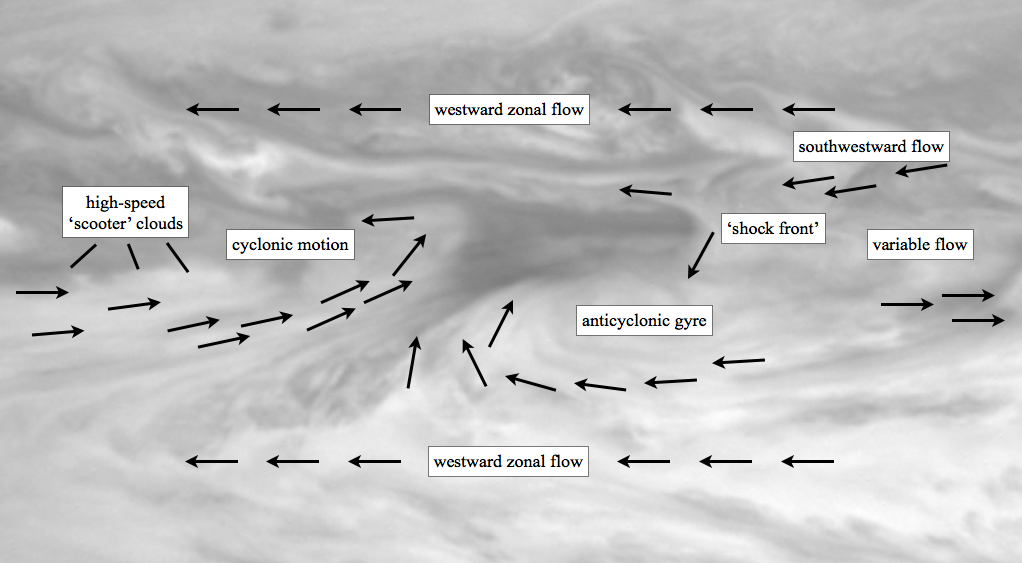}
  \caption[Hot spot Dynamics Sketch]{
    \label{Figure: hotspot_dynamics_sketch}
    Sketch illustrating typical flow patterns throughout the entire period of our data set in the vicinity of a Jovian hot spot from an overhead observer. All vectors are drawn in a frame of reference traveling with the hot spots. Vectors only represent prevailing direction of wind flow, and not magnitude.
    }
\end{figure}

Figure \ref{Figure: hotspot_dynamics_sketch} illustrates dynamical patterns consistently observed across several hot spots from the perspective of an observer traveling at the hot spot drift speed of 103 m s$^{-1}$ eastward throughout the entire observational period. Note that these typical dynamics are usually short-lived, as constant interactions with storms from surrounding latitudes disrupt this baseline state. As storm interactions diminish, however, the dynamics typically return to this state, though some characteristics depicted in Figure \ref{Figure: hotspot_dynamics_sketch} are not always present. The zonal flow to the north and south, anticyclonic gyres, and high-speed 'scooter' clouds are the most common dynamical features noted during the period of study, and are hallmark features of the baseline state.

The 'normal' dynamics surrounding a hot spot begin with jets bounding these features at their north and south that flow to the west relative to the hot spots. The jet to the north also transports clouds and storms adequately southward so that they constantly buffet against a hot spot, shaping the appearance of its northeast corner. As discussed in the previous section, hot spots tend to have a sharply defined eastern edge, and its interaction with cloud features in surrounding latitudes from the perspective of an observer traveling with the hot spots produces the impression of a shock front. Some incoming cloud material diverts either to the north and west, where it simply advects westward with slower flow. The remaining material diverts to the south, where these clouds are possibly influenced by the faster 7$^{\circ}$N jet. These clouds either become entrained in the anticyclonic gyre or stagnate in a variable flow area directly to the east of the hot spot. 

This area directly east of the hot spot is particularly complex, as it does not exhibit any characteristic flow. From the time-lapse movies, the presiding flow appears subject to whatever external influences (passing storms, gyres, or plumes) are dominant. Because this area also lies between regions of stronger westward flow to the north, and eastward flow to the south from the anticyclonic gyre or the western portion of an equatorial plume, strong latitudinal shear should be present. The morphology of the cloud features, however, do not suggest the presence of wind shear.  

\begin{figure}[htbp]
  \centering
  \includegraphics[keepaspectratio=true, width=5.5in]{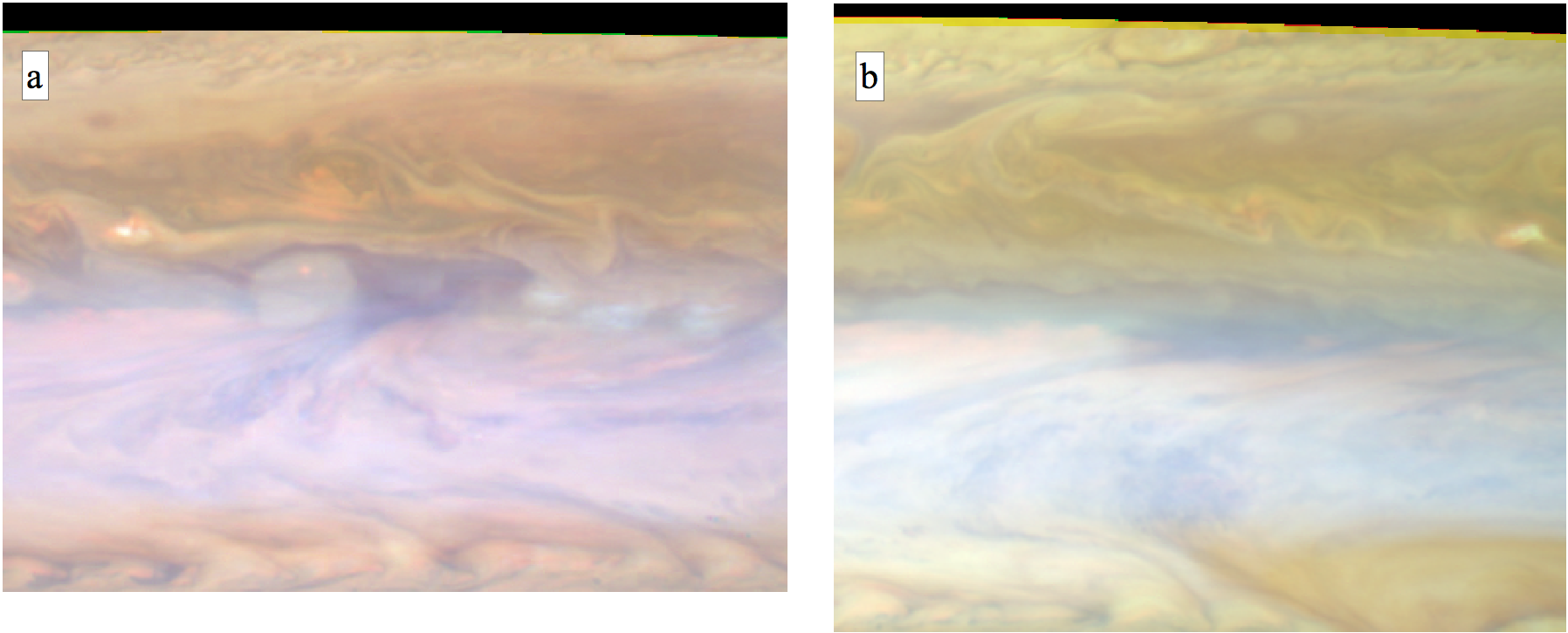}
  \caption[Hot spot false-color]{
    \label{Figure: hotspot_rgb}
    \emph{Cassini} false-color image of a hot spot and its surroundings. The red channel represents CB2 near-infrared continuum data, the green channel represents MT2 (727 nm) methane band, and the blue channel represents MT3 (889 nm) methane band. (a) The reddish tint of the plume to the hot spot's southwest and gyre to the hot spot's south indicate that both are located in the lower troposphere, whereas the bluish tint of clouds east of the hot spot indicate that they are located in the upper troposphere and stratosphere. Note the discrete deep cloud spot beneath a higher cloud directly west of the hot spot, the cloud/haze feature located at higher altitudes directly over a hot spot, and the general haze pervasive throughout the equatorial latitudes. (b) In contrast, this hot spot exhibits a cloud within a hot spot located at deeper latitudes from its reddish tint. Time-lapse movies indicate that this cloud likely originated from the plume situated to the west of a hot spot. Note the lack of upper haze above the South Equatorial Disturbance at the bottom right.
    }
\end{figure}

Instead, diagnosis of the dynamics occurring in this region is complex because of flow occurring at different altitudes overlain on one another. Figure \ref{Figure: hotspot_rgb} is a two-image panel of false-color images where the red channel represents brightness from the near-infrared continuum (CB2), green represents the 727 nm methane band (MT2), and blue represents the 889 nm methane band (MT3). As a result, bluer features are located in the upper troposphere and stratosphere, while redder features are located in the deeper troposphere. Therefore, to the east of the hot spot, we are likely observing cloud features at different altitudes overlain on one another as viewed from above. 

Two distinct colors for cloud features are present in Figure \ref{Figure: hotspot_rgb}a to the east and south of a hot spot. Clouds represented as lighter blue in the figure originate from the north and east, as the jet to the north advects them directly towards interaction with the eastern front of the hot spot. These colors suggest that they are primarily in the upper troposphere, though perhaps somewhat vertically extended given their more substantial white color. Contrasting with these features are redder cloud streaks residing within the anticyclonic gyre, indicating that these features are located in the lower troposphere, presumably at the condensation level of ammonia in Jupiter's atmosphere ($\sim$1 bar). Overall, advancing cloud material from the north of a hot spot appears to obscure motions occurring underneath them, complicating the measurement of wind outflow away from a hot spot, and the general dynamics occurring east of the feature.

The atmospheric structure revealed by Figure \ref{Figure: hotspot_rgb} also helps constrain dynamics at other locations. In both panels of Figure \ref{Figure: hotspot_rgb}, a wide swath of upper tropospheric haze is present in a zonal band approximately straddling the equator up to the southern portions of a hot spot. Haze patches occasionally present over a hot spot indicate that any strong vertical motion is confined to deeper altitudes \citep{Banfield98}. This confinement is consistent with the absence of prominent features in thermal infrared maps at the hot spot and plume latitudes from \emph{Cassini} CIRS observations \citep{Li06Icarus}. To the west of a hot spot, perhaps under-recognized is the observation made by \citet{Vasavada98} of cyclonic rotation in clouds to the west of a hot spot. We observe consistent cyclonic rotation in these areas west of most hot spots in our data sets. In certain areas and times, clouds traverse a complete cyclonic circuit to the west of the hot spot before advecting to adjacent latitudes and integrating into the normal eastward or westward flow of nearby jet streams. From figure \ref{Figure: hotspot_rgb}a, these clouds are located at higher altitudes and most likely represent an independent dynamic regime presumably separate from the flow patterns directly controlling a hot spot.

Our movies also confirm what was noted previously from \emph{Galileo} observations by \citet{Vasavada98}: the northeastward flow near the southwestern corner of a hot spot is partly associated with anticyclonic gyres present to the southeast of hot spots. Furthermore, this flow pattern is consistent to hot spots at all locations and is not isolated to one or a few locations. \citet{Garcia-Melendo11}, however, recently reported convergent flow in only 2 out of 9 image pairs depicting hot spot flow from their analysis of \emph{Cassini} and HST imagery, and suggested that the presence of convergent flow may be dependent on hot spot morphology. From our time-lapse animations, convergent flow with anticyclonic gyres nearby is clearly present at all mature, well-formed hot spots, but such flow is not present or not visible for hot spots that are smaller, migratory, or ephemeral. Therefore, our analysis supports this suggestion, though we cannot determine if a causal relationship exists between morphology and dynamics.  

\begin{figure}[htbp]
  \centering
  \includegraphics[keepaspectratio=true, width=6in]{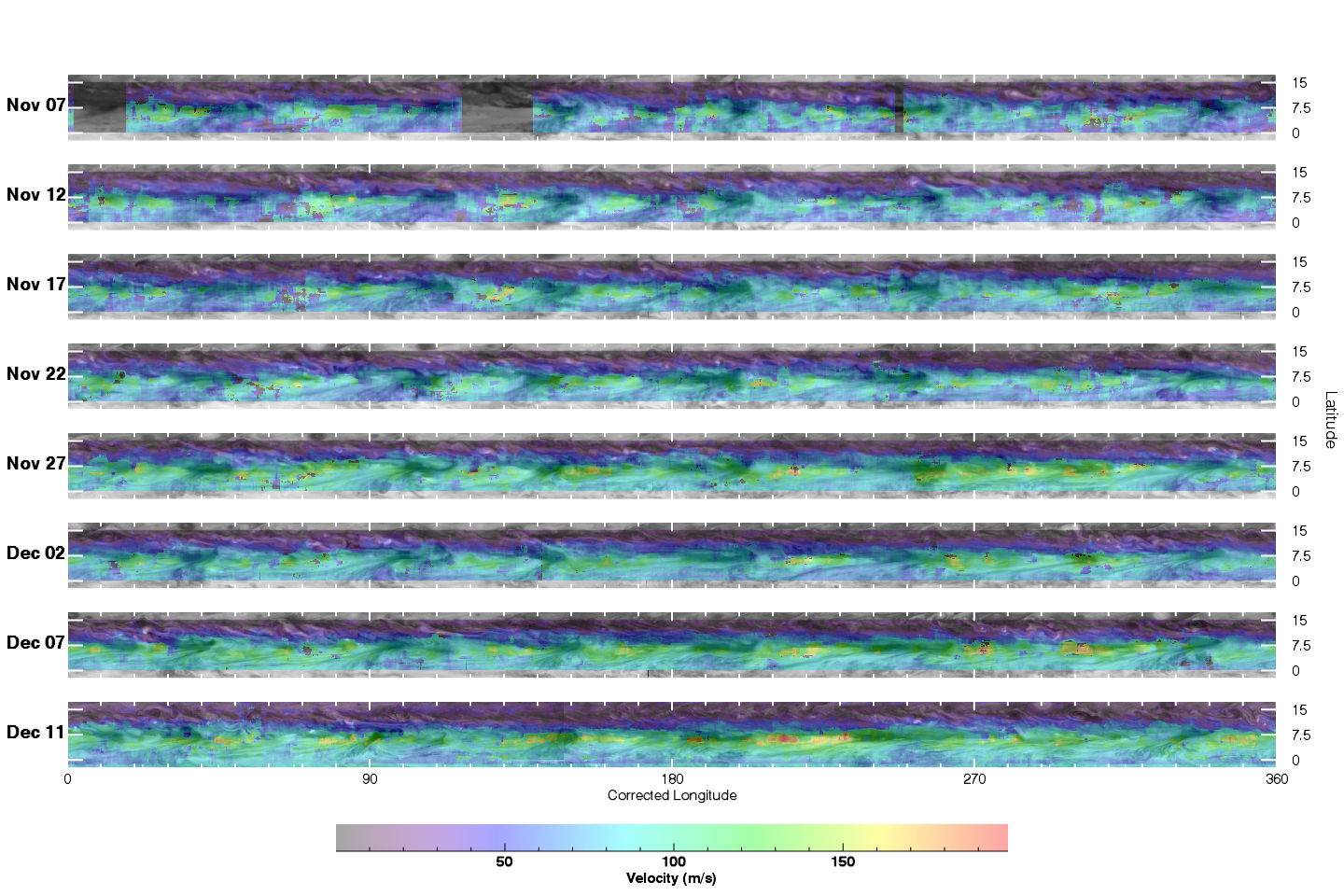}
  \caption[Hot Spot Velocity Maps vs. Time]{
    \label{Figure: hotspot_velocityvstime}
    A time-series of color velocity maps overlain on near-infrared continuum mosaics. The velocities are measured from a System III reference frame, whereas each map and mosaic are longitudinally shifted to a reference frame traveling east at 103 m s$^{-1}$, the propagation speed of the hot spots.
    }
    \label{lastfig}
\end{figure}

\begin{figure}[htbp]
  \centering
  \includegraphics[keepaspectratio=true, width=5.25in]{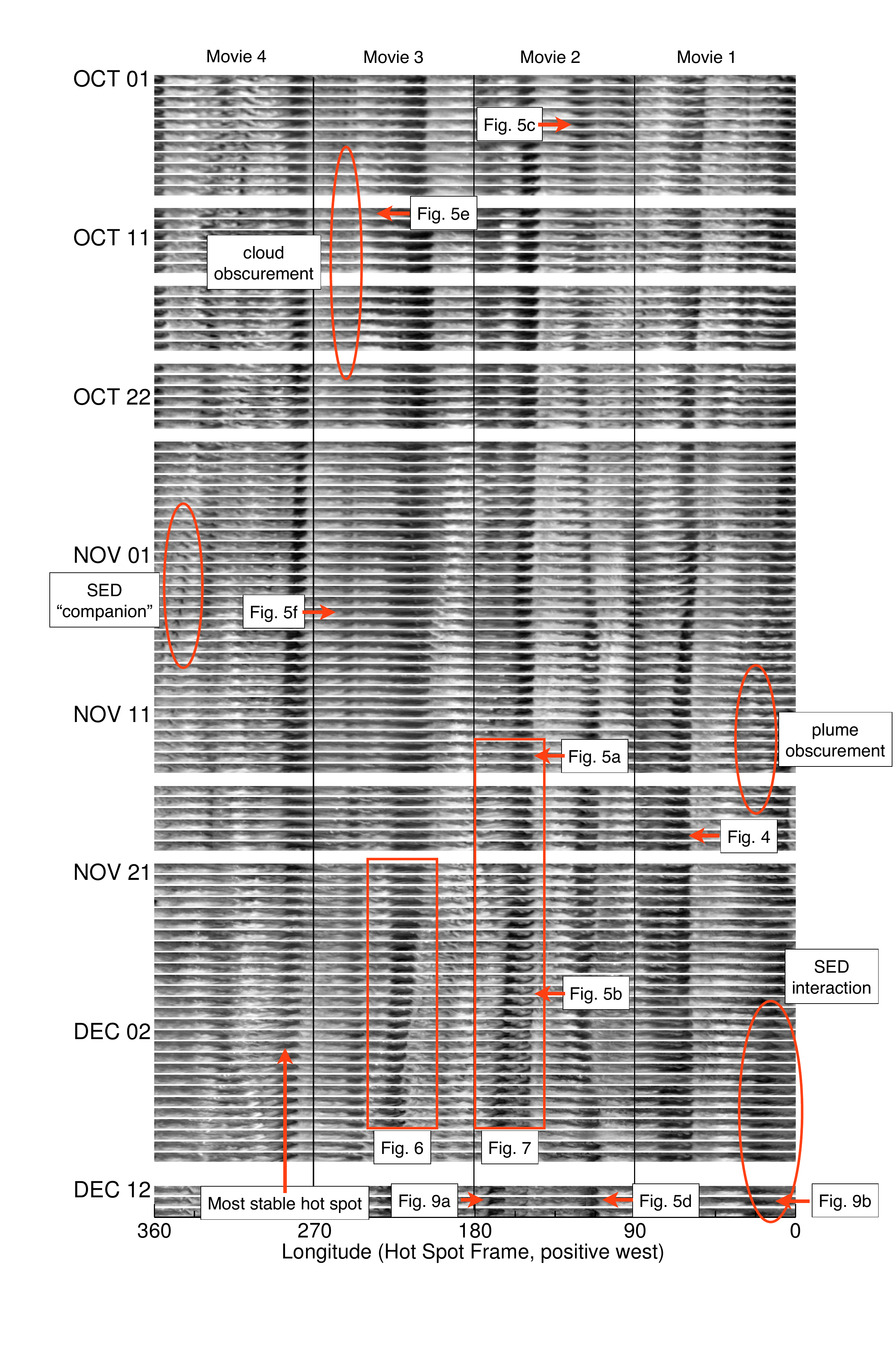}
  \caption[Supplementary Figure 1]{
    \label{Figure: suppl_fig1}
    An annotated version of Figure 1 from the manuscript, marking locations and times of various events referenced throughout the text and other figures.
    }
    \label{lastfig}
\end{figure}

Also reported in \citet{Choi11} and \citet{Garcia-Melendo11} are observations of very fast flow within plumes or bright cloud areas to the west of hot spots. Our inspection of the time-lapse movies reveal that these measurements stem from compact bright, white `scooter' clouds that rapidly traverse across the background plume. Figure \ref{Figure: hotspot_velocityvstime} shows velocity magnitude mapped as color overlain on a time series of near-infrared image mosaics. This figure demonstrates that the location of high velocity measurements correlate well with bright white cloud tops within equatorial plumes.  These 'scooter' clouds typically travel at speeds 150 m s$^{-1}$ and above, with certain isolated measurements suggesting speeds approaching 200 m s$^{-1}$. Figure \ref{Figure: hotspot_velocityvstime} and the time-lapse movies also show that these clouds are common within plumes, and are consistently present throughout the observation period.

Typically, these clouds disappear and presumably evaporate as they enter the dark area of a hot spot. In some cases, however, small portions remain visible within the hot spot, as seen in Figure \ref{Figure: hotspot_rgb}b. These clouds are also those reported by \citet{Li06}, who described their fast velocity relative to standard zonal wind profiles of Jupiter at that latitude, and who surmised that these clouds were deep water clouds in the lower troposphere. However, the velocities reported by \citet{Li06} are similar to the velocities present in measurements of 'scooter' clouds, and time-lapse movies suggest that the deep cloud seen within the hot spot pictured in Figure \ref{Figure: hotspot_rgb}b originated as a scooter cloud from an equatorial plume. 

\subsection{Plume Morphology}

Plumes distinguish themselves from their surroundings as an area of bright clouds, with even brighter small scooter clouds traveling rapidly within the plumes. In general, the brightest cloud tops of the plumes occur along their northern edge; this edge is typically aligned with the central latitudinal axis of the hot spots. The fanning out of clouds from a plume head that produce the comet-like tail tends to angle towards the south and southwest. This is probably associated with the meridional wind shear present in the jet stream as eastward velocity decreases towards the equator. The strength and position of the anticyclonic gyre located to the southeast of a hot spot may also be a controlling factor for the particular angle that a plume fan cloud adopts. As a gyre sometimes migrates to its west, its leading edge appears to sculpt a plume's southeastern boundary. A curved, dark streak dividing a plume from a hot spot also forms, perhaps caused by the approaching gyre.

Differences in plume morphology are present when considering the density of scooter clouds within a plume. Nearly all plumes exhibit these clouds, but in plumes with a high density of scooter clouds, there is typically no plume head or 'focal point' for the plume located near the western boundary of a hot spot. In these cases, scooter clouds constantly spawn within this wedge/plume area, and then flow rapidly eastward, ultimately flowing into the hotspot, where they are absorbed or otherwise disappear from view. Furthermore, the scooter clouds are brighter than the surrounding plume, with the plume itself not particularly distinguishable from its surroundings. These contrasting plume forms can be located near one another, as in one case, two plumes on the opposite sides of a hot spot exhibited striking disparities. One was a collection of discrete, bright clouds comprised one `plume,' whereas the other plume appeared distinguished and bright, with reduced instances of isolated 'scooter' clouds. Both plumes maintained their position relative to one another and the hot spot, though one plume appeared to slowly migrate to the east. 

Occasional interactions between hot spots and plumes typically favor the hot spot. In one instance, growth of a hot spot seemingly causes absorption of a plume located to its west. During this encounter, the plume's morphology does not change appreciably, instead forming an interesting juxtaposition of a bright plume adjacent to a dark hot spot. Scooter clouds within the plume continue to enter the hot spot (and disappear from view) at its southwestern quadrant as the hot spot encroaches. However, when hot spots weaken or otherwise shrink in longitudinal width, adjacent plumes start to re-appear into view or expand into the former space occupied by the hot spot.

Storms originating from the North Equatorial Belt not only interact with hot spots in shaping their northeastern edges, but also influence plumes. As clouds from passing storms travel southward and encounter the hot spots and plumes, these events appear to trigger brightening of cloud spots within the plumes. Thus, these incoming waves may act as catalysts for convection or outbursts of cloud formation. The degree of interaction between a plume and the incoming material is also variable. More common is a glancing interaction, where the incoming storm triggers bright, but ephemeral, compact clouds within a plume. At certain other times, however, the avenue of incoming cloud material appears to linger within a plume, somewhat brightening the plume overall as it generates more clouds and perhaps acting as a stimulus for increased plume activity. These interactions appear to be consistent throughout the era of spacecraft exploration of Jupiter, as similar activity was noted by \citet{Beebe89} from \emph{Voyager} data.

\section{Discussion}

In summary, our notable results from our analysis of \emph{Cassini} observations are the following: 

\begin{itemize}
\item Hot spots can experience significant changes in size and shape over daily-to-weekly timescales. However, the more prominent ones maintain a consistent propagation speed and spacing throughout our two-month data set, implying that the equatorial Rossby wave theorized to control their formation and behavior is stable on monthly timescales.
\item Strong anticyclonic gyres are located to the south and southeast of prominent, mature hot spots, confirming previous analysis from \emph{Galileo} observations. In addition, cyclonic gyre circulation is occasionally present to the west of hot spots. Previous dynamical models of Jupiter's equatorial atmosphere also exhibit gyre circulations, as discussed later within this section.
\item Passing vortices to the north and the South Equatorial Disturbance to the south are associated with transformations in the size and shape of hot spots, as well as splitting events. These splitting events imply a degree of non-linearity in the wave dynamics associated with hot spots and plumes. 
\item Bright, compact ``scooter'' clouds travel at speeds faster than the background propagation speed of the hot spots and plumes. These clouds are the source of positive velocity perturbations relative to the zonal wind profile at 7$^{\circ}$N.  
\end{itemize}

The observations of scooter clouds raise questions over whether these clouds travel at anomalously high speeds relative to published zonal wind profiles \citep{Limaye86, Garcia-Melendo01, Porco03} or whether these clouds represent the flow of the background jet stream. If a planetary wave is the mechanism responsible for generating the hot spot and plume features, the speed of hot spot propagation should be interpreted as the phase speed of the wave in System III longitude. Because this wave is a Rossby wave propagating westward relative to the zonal jet, the true speed of the zonal jet at the hot spot latitude is higher than the drift speed of the hot spots in System III. Most latitudinal zonal wind profiles of Jupiter, however, base their measurements on correlating latitudinal strips encompassing all available observed longitudes. For latitudes containing the hot spots and plumes, this method cannot avoid measuring the propagation speed of these dominant features. 

Therefore, some caution must be considered when interpreting these zonal wind profiles, particularly at these latitudes. \citet{Beebe96} first raised cautions regarding published reports basing wind measurements on large equatorial weather features, and demonstrated that the small cloud features near 7$^{\circ}$N traveled at $\sim$150 m s$^{-1}$. \citet{AsayDavis11} also raised this concern, noting that previous measurements of zonal winds were largely measuring the propagation speed of the hot spots. From a modeling perspective, the simulated hot spots in \citet{Showman00} tended to travel westward relative to the jet at $\sim$70 m s$^{-1}$, implying a jet speed of $\sim$170 m s$^{-1}$, consistent with the speed of the scooter clouds in our study. Collectively, these studies, along with \citet{Garcia-Melendo11}, suggest a higher wind speed for the jet near 7$^{\circ}$N, resulting in a latitudinally symmetric equatorial jet profile; our measured velocities for the scooter clouds also support this interpretation. 

If the hot spots and plumes are manifestations of a trapped equatorial Rossby wave, zonal wind measurements that do not account for these phenomena will undervalue the jet's inherent speed, because the hot spots and plumes are traveling westward relative to the jet (but eastward relative to an inertial frame). Analysis of the dispersion relationship of this planetary wave by \citet{Arregi06} suggests that the background zonal flow is closer to $\sim$140 m s$^{-1}$, which would render the equatorial `cusp' feature of Jupiter's latitudinal zonal wind profile somewhat more symmetrical about the equator. 

Note that the \emph{Galileo} probe results \citep{Atkinson98} directly measured winds of 150 m s$^{-1}$ at about the 3 bar level, perhaps indicating that the jet deflects downward to at least this pressure level within an equatorial hot spot. Furthermore, the probe results reported 170 m s$^{-1}$ winds at 4 bars that remained fairly constant down to 20 bars. Therefore, the revised zonal wind at 7$^{\circ}$ implies reduced vertical wind shear than previous estimates, which may be affected by the local meteorology of the hot spots. These reduced vertical shears match predictions made by \citet{Showman00} of little to no vertical shear in the middle of a hot spot, and that this shear is a local consequence of the hot spot, and not a systematic characteristic of the jet as a whole. Subsequent work by \citet{Sanchez-Lavega08} examining simulations of plume eruptions within the strong North Tropical Jet on Jupiter also suggests minimal vertical wind shear at other latitudes.

A related issue regarding this equatorial planetary wave and its relation to the overall dynamics occurring in the vicinity of the hot spot concerns the re-emergence of the flow once it travels through a hot spot. No systematic flow pattern is present in areas immediately east of a hot spot, where flow should emerge from depth, triggering cloud condensation and plume formation. This is partially caused by higher altitude haze layers obscuring deeper altitudes, preventing suitable measurements at the levels of interest. However, given the presence of anti-cyclonic gyres consistently present to the south and southeast of hot spots, it is plausible that air re-emerges and flows out of a hot spot somewhere to its southeast, matching results reported by \citet{Showman00}. The emergence of flow out of hot spots at these locations may be a consequence of meridional perturbations of this equatorial wave, causing the southward tilt of the plume cloud, the southwest-to-northeast flow from plume to hot spot, and the positioning of a plume to the south of a hot spot. However, observations do not suggest a latitudinal perturbation of the hot spots and plumes themselves, similar to chevron behavior influenced by a Rossby wave within an inertia-gravity wave \citep{Simon-Miller12}. Instead, the latitudinal position of these features is steady as they drift westward relative to the background jet. 

Note that the anticyclonic gyres could be part of an alternating, von Karman-like pattern of gyres at this latitude. However, the weak Coriolis forces present at low latitudes impede a classical vortex's formation and stability near the equator, as vortices tend to destabilize via Rossby wave radiation. These gyres are instead likely a manifestation of a Rossby wave, which is also comprised of alternating regions of cyclonic and anti-cyclonic vorticity. Northwest of these gyres and wedged in between hot spots to their west and plumes to their north were areas of cyclonic circulation, with cloud features embedded in these areas occasionally approaching the hot spots from the west, diverting northward along the boundary with the hot spot, and then advecting westward along the boundary with the North Equatorial Belt. Occasionally these features trace complete cyclonic circuits in these areas, and are likely driven by the meridional perturbations of the equatorial wave and the surrounding flow. Based on Figure \ref{Figure: hotspot_rgb}, differences in altitude also complicate this model, as the cyclonic motions observed west of the hot spot originate from upper atmospheric hazes, whereas the anticyclonic gyre is likely located deeper in the troposphere within the main ammonia cloud deck. The vertical extent of these circulations remains an open question. 

Interestingly, the nominal dynamical model by \citet{Friedson05} that assumed vertical wind shear matching the Doppler wind profile of the \emph{Galileo} probe predicts paired cyclonic-anticyclonic gyres crudely similar to the current observations. When assuming a constant vertical wind profile, however, only a single gyre with the hot spot at its center is present. This is also crudely similar to observations in that the observed anticyclonic gyre is one of the more dominant dynamical features affecting a hot spot, though the hot spot is generally located to the north or northwest of the gyre. \citet{Friedson05} remarks that the model is very sensitive to the vertical profile of the zonal wind. The true vertical wind shear likely lies in between the tested cases, providing motivation for a re-examination of this model to see if its results improve its match with observations. 

As discussed earlier, the original proposal set forth by \citet{Stoker86} suggested that bright spots at the ``head'' of a plume could be a site of active convection, with the ``tail'' of the plume representing the top of a thunderstorm anvil cloud stretched by the ambient jet stream. However, these plume ``heads'' have been largely absent since the \emph{Voyager} observations. Furthermore, the lack of observed lightning \citep{Gierasch00}, combined with only a few plume heads appearing bright white and suggestive of convection \citep{Rogers95}, has advanced the alternative planetary equatorial wave theory as a more favorable explanation for the plume phenomena. In addition, the rapid flow of scooter clouds towards a plume ``head'' observed in the present study does not reconcile with expected outflow from a proposed site of vigorous convection. Instead, the mostly fixed position of plumes relative to the hot spots throughout the observation period lends credence to the theory that a single planetary wave controls their drift, and that the upwelling portion of this wave produces plumes. 

The majority of detections of water ice spectral signatures on Jupiter using \emph{Voyager} IRIS data were located at 10$^{\circ}$N \citep{Simon-Miller00}. Inspection of the time-lapse footage reveals numerous examples of convective events in the boundary between the hot spots and the North Equatorial Belt. During these sporadic events, concentrated areas of very bright clouds develop rapidly and experience impressive growth. It is plausible that these convective events may drive water vapor from its normal condensation level ($\sim$4 bar) to altitudes above the main ammonia cloud deck where it can detected spectrally \citep{Banfield98, Gierasch00}. Therefore, when considering the depletion of ammonia observed within plumes \citep{Showman05}, the spectral detections of water ice likely originates  from these convective outburst events. The water ice could establish some residence aloft, where ongoing convection continually supplies water ice at these high altitudes. In addition, \citet{Baines02} detected spectral features of solid ammonia ice between 2$^{\circ}$ and 7$^{\circ}$N, just south of the hot spots, and consistent with the implication of active uplift and/or convection throughout the region. A more recent observation from the \emph{New Horizons} flyby \citep{Reuter07} also associates regions of vigorous upwelling with detections of ammonia ice, though their primary detection was located in temperate latitudes.

Finally, our analysis has demonstrated the utility of time-lapse imagery for illustrating the diversity of fluid behaviors and interactions present in Jupiter's atmosphere. However, while intermittent visits by spacecraft provide high-fidelity measurements of various aspects of planetary systems, they have become somewhat restrictive in enabling long-term studies of meteorological phenomena on other planets. One solution is a low-cost, dedicated monitoring telescope for planetary science in Earth orbit \citep{Noll12} set to constantly observe Jupiter. The increased spatial resolution, frequent cadence of observations, and overall mission length all combine for advantageous data collection of time-variable meteorological phenomena in the outer planets. Detailed analysis of the data carries the potential for breakthroughs in our understanding of giant planet atmospheric dynamics.

\section{Acknowledgements}
We thank two anonymous reviewers for their comments that strengthened this manuscript. This research was supported by a NASA Jupiter Data Analysis Program grant, \#NNX09AD98G, and by an appointment to the NASA Postdoctoral Program at Goddard Space Flight Center, administered by Oak Ridge Associated Universities through a contract with NASA.

\label{lastpage}

\bibliographystyle{model2-names} 
\bibliography{hotspot-bib}

\end{document}